\documentclass[twocolumn,showpacs,preprintnumbers,amsmath,amssymb]{revtex4}
\usepackage{graphicx}
\usepackage{dcolumn}
\usepackage{bm}
\input{epsf.tex}

\begin{document}

\title{Evidence for a Narrow N$^*$(1685) Resonance  
in Quasifree Compton Scattering on the Neutron}

\author{V.~Kuznetsov$^{1,2}$}
\author{M. V.~Polyakov$^{3,4}$}
\author{V.Bellini$^{5,6}$}
\author{T.~Boiko$^7$}
\author{S.~Chebotaryov$^1$}
\author{H.-S.Dho$^{1}$}
\author{G.Gervino$^{8,9}$}
\author{F.Ghio$^{10,11}$}
\author{A.Giusa$^{5,6}$}
\author{A.~Kim$^{1,12}$}
\author{W.~Kim$^{1}$}
\author{F.Mammoliti$^{5,6}$}
\author{E.~Milman$^1$}
\author{A.~Ni$^1$}
\author{I.A.~Perevalova$^{13}$}
\author{C.Randieri$^{5,6}$}
\author{G.Russo$^{5,6}$}
\author{M.L.Sperduto$^{5,6}$}
\author{C.M.Sutera$^{5}$}
\author{A.N.~Vall$^{13}$}

\affiliation{$^1$Kyungpook National University, 702-701, Daegu,Republic of Korea}
\affiliation{$^2$Institute for Nuclear Research, 117312, Moscow, Russia}
\affiliation{$^3$Institute f\"ur Theoretische Physik II, Ruhr-Universit\"at
Bochum, D - 44780 Bochum, Germany,}
\affiliation{$^4$Petersburg Nuclear Physics Institute, Gatchina, 188300, St. Petersburg, Russia,}
\affiliation{$^{5}$INFN - Sezione di Catania, via Santa Sofia 64, I-95123 Catania, Italy}
\affiliation{$^{6}$Dipartimento di Fisica ed Astronomia, Universit\'a di Catania,
via Santa Sofia 64, I-95123 Catania, Italy} 
\affiliation{$^7$Belarussian State University, 220030, Minsk,
Republic of Belarus}
\affiliation{$^{8}$Dipartimento di Fisica Sperimentale, Universit\'a di Torino, 
via P.Giuria, I-00125 Torino, Italy}
\affiliation{$^{9}$ INFN - Sezione di Torino, I-10125 Torino, Italy}
\affiliation{$^{10}$INFN - Sezione di Roma, piazzale Aldo Moro 2, I-00185 Roma, Italy}
\affiliation{$^{11}$Instituto Superiore di Sanit\'a, viale Regina Elena 299, I-00161 Roma, Italy}
\affiliation{$^{12}$Thomas Jefferson National Accelerator Facility, Jefferson Av., 23606 VA, USA}
\affiliation{$^{13}$ Physics Department, Irkutsk
State University, Karl Marx str. 1, 664003, Irkutsk, Russia }
\date{\today}

\begin{abstract}

The first study of quasi-free Compton scattering on the neutron in
the energy range of $E_{\gamma}=0.75 - 1.5$ GeV is presented. The
data reveals a narrow peak at $W\sim 1.685$ GeV. This result,
being considered in conjunction with the recent evidence for a narrow structure at
$W\sim 1.68$~GeV in the $\eta$ photoproduction on the neutron, suggests
the existence of a new nucleon resonance with unusual properties:
the mass $M\sim 1.685$~GeV, the narrow width $\Gamma \leq 30$~MeV,
and the much stronger photoexcitation on the neutron than on the
proton.

\end{abstract}

\maketitle

Many properties of known baryons were transparently
explained by the constituent quark model(CQM)~\cite{isg} that treats
baryons as bound system of three valence quarks in the ground or
excited state. Some
baryon properties remain a mystery:
almost half of the CQM-predicted nucleon and $\Delta$
resonances~\cite{cap} still escape the reliable experimental
identification~\cite{pdg} (so-called ``missing resonances").

The chiral quark soliton model ($\chi$QSM) is an alternative view
of baryons which are treated as space/flavor rotational
excitations of a classical object - a chiral mean-field. $\chi$QSM
predicts the lowest-mass multiplets of baryons to be the $1/2^+$
octet and $3/2^+$ decuplet - exactly as CQM does. The $\chi$QSM
predictions for higher multiplets are different from
CQM~\cite{dia}.

Thus, the experimental study of baryon resonances provides
benchmark information for the development of theoretical models
and for finding relations between them.

In this context the possible observation of a new narrow resonance
$N^*(1685)$ is of
potential importance. Recently, four groups - GRAAL~\cite{gra1},
CBELSA/TAPS\cite{kru}, LNS~\cite{kas}, and Crystal
Ball/TAPS~\cite{wert} - reported evidence for a narrow structure
at $W\sim 1.68$~GeV in the $\eta$ photoproduction on the neutron.
The structure was observed as a bump in the quasi-free cross
section and as  a peak in the invariant-mass spectrum of the
final-state $\eta$ and the neutron $M(\eta n)$
~\cite{gra1,kru,wert}. The width of the bump in the quasi-free
cross section is close to the smearing caused by Fermi motion of the
target neutron bound in a deuteron target~\cite{gra1}. The
width of the peaks observed in the $M(\eta,n)$ spectra is close
the instrumental resolution of the corresponding
experiments~\cite{gra1,kru,wert}.

Furthermore, a sharp resonant structure at $W\sim 1.685$~GeV was
found in the GRAAL beam asymmetry data for the $\eta$
photoproduction on the free proton ~\cite{acta,jetp}(see also ~\cite{an1}).
Such structure is not (or poorly) seen in the $\gamma p \to \eta p$
cross section~\cite{crede}. Any resonance whose photoexcitation on
the proton is suppressed by any reason, may manifest itself in
polarization observables due to interference effects.

In Refs.~\cite{gra1,az,tia,kim,acta,jetp}, the combination of the
experimental findings was interpreted as a possible signal of a nucleon
resonance with unusual properties: the mass near $M\sim 1.68$~GeV, the
narrow width, and the strong photoexcitation on the neutron.
Alternatively, the authors of  Refs.~\cite{ani,skl} explained
the bump in the quasi-free $\gamma n \to \eta n$ cross section 
in terms of the interference of well-known resonances or as the virtual
sub-threshold $K \Lambda$ and $K \Sigma$ photoproduction~\cite{dor}.

If the narrow $N^*(1685)$ does really exist, it can be seen
not only in the $\eta$ photoproduction
but also in other reactions on the neutron, e.g. Compton scattering
or the $\pi^0$ photoproduction.
On the contrary, the narrow bump cannot be generated by the
interference of wide resonances in these reactions, as they
receive contributions of different (from the $\eta$
photoproduction) resonances.

In this paper, we present the first measurement of Compton
scattering on the neutron at the photon energies of
$E_{\gamma}=0.75 - 1.5$~GeV ($W \sim 1.5 - 1.9$~GeV) focusing on
the search for the signal of $N^*(1685)$. Simultaneously, we
investigate the photoproduction of neutral pions on the neutron
and the isospin-mirrored reactions $\gamma p \to \gamma p$ and
$\gamma p \to \pi^0 p$.

The existing data base for Compton scattering and $\pi^0$
production on the neutron is scarce. The available data for
Compton scattering is limited to lower energies $E_{\gamma} \leq
400$~MeV~\cite{rose,kolb,kost}. There is no published data for
$\gamma n \to \pi^0 n$ cross section in the domain of $W\sim 1.6
- 1.7$~GeV. New preliminary data on the $\gamma n \to \pi^0 n$
cross section were presented in \cite{kru1,shim}.

\begin{figure}
\vspace*{0.4cm}
\centerline{\epsfverbosetrue\epsfxsize=7.9cm\epsfysize=7.cm\epsfbox{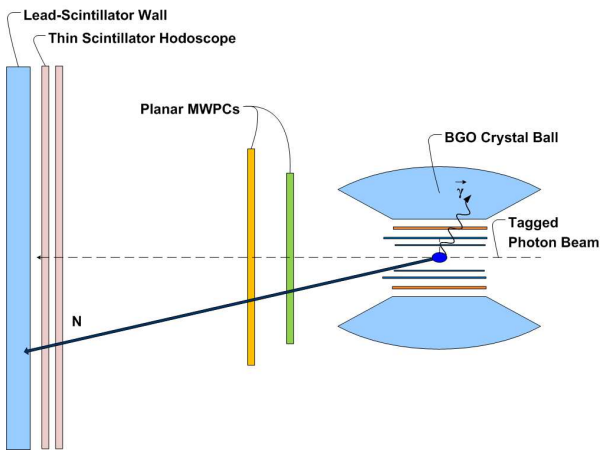}}
\caption{Schematic view of the GRAAL detector.} \vspace*{-0.3cm}
\label{fig:kin} \vspace{-0.3cm}
\end{figure}

The data was collected at the GRAAL
facility~\cite{pi0}. The GRAAL polarized and tagged photon beam is
produced by backscattering of laser light on 6.04~GeV electrons
circulating in the storage ring of the ESRF (Grenoble, France).
The $4\pi$ detector (Fig.~\ref{fig:kin}) is designed for the
detection of neutral and charged particles. It is composed of a
cylindrically symmetrical central part, for the detection of the
particles emitted at $\theta_{lab}=25 - 155^{\circ}$ with respect
to a beam axis, and of a forward part for the detection of the
particles emitted at $\theta_{lab}\leq 25^{\circ}$.

The central part consists of two coaxial cylindrical wire
chambers, a 5~mm thick plastic scintillator barrel, which provides
$\Delta E$ information for particle identification,  and a BGO
ball made of 480 crystals each of 21 radiation length.
The energy resolution for the detection of photons at $1$ GeV is
3\%(FWHM).

 The forward part consists of two planar multiwire chambers, which provide
 tracking angular resolution of $\sim$0.5$^{\circ}$ for charged particles,
 and a double hodoscope wall
 made of two layers of 3 cm thick plastic scintillator
 bars covering an area of 3x3 m$^{2}$ and located 3 m away from the target.
 The hodoscope wall is followed by the TOF lead-scintillator
 wall which is an assembly of 16 modules covering the same area as the
 hodoscope wall.  Each module is a composition of four
 300x19x4 cm$^3$ scintillator bars separated by 3 layers of 3~mm thick lead
 converter. The wall provides the detection of photons, neutrons and charged
 particles with an angular resolution of 3$^{\circ}$ and $\Delta$E information. 
 The TOF resolution is 600 ps (FWHM) for
 charged particles and 700-800 ps for neutrons.
 The estimated efficiency of the detection of photons and neutrons is about 95 and 22\%
 respectively. The particle identification (photons, neutrons, protons or charged pions)
 in the forward assembly is achieved by means of coincidence (anticoincidence)
 of the corresponding signals in the lead-scintillator wall and the preceding
 planar chambers and the hodoscope wall, and using $\Delta$E-TOF relations.
 Momenta of the charged particles and neutrons can be reconstructed from the
 measured TOF and angular quantities.

Both $d(\gamma,\gamma^{\prime}n)p$ and $d(\gamma,
\gamma^{\prime}p)n$ reactions were measured simultaneously  in the
kinematics that emphasize the quasi-free reaction. Scattered
photons were  detected in the BGO crystal ball~\cite{bgo}. Recoil
neutrons and protons emitted at $\Theta_{lab} = 3 - 23^{\circ}$
were detected in the assembly of the forward detectors
(Fig.~\ref{fig:kin}).

As the first step, the identification of $\gamma N$ final states
was achieved using the criterion of coplanarity, cuts on the
neutron(proton) and photon missing masses, and comparing
the measured TOF and the angle of the recoil nucleon with the same quantities
calculated assuming  the $\gamma N \to \gamma N$ reaction.
The sample of the selected events was still contaminated by the
events from the $\pi^0$ photoproduction. The $\pi^0$ cross section
is about two orders of magnitude larger than that of Compton
scattering.

At the second step, two types of the $\pi^0$ background
were taken into consideration:\\
i) Symmetric $\pi^0\to 2\gamma$ decays. The pion decays in two
photons of nearly equal energies. Being emitted in a narrow cone
along the pion trajectory,
such photons imitate a single-photon hit in the BGO ball;\\
ii) Asymmetric $\pi^0\to 2\gamma$ decays. One of the photons takes
the main part of the pion energy. It is emitted nearly along the
pion trajectory. The second photon is soft and is emitted into a
backward hemisphere relative to the pion track. Its energy depends
on the pion energy and may be as low as $6 - 10 $~MeV.

The symmetric events were efficiently rejected by analyzing the
distribution of energies deposited in crystals attributed to the
corresponding cluster in the BGO ball. The efficiency of this
rejection was verified in simulations and found to be $99\%$.

The asymmetric $\pi^0\to 2 \gamma$ decays present the major
problem. The GRAAL detector provides the low-threshold ($5$~MeV)
detection of photons in the nearly $4\pi$ solid angle. If one
(high-energy) photon is emitted at backward angles
$\Theta_{lab}=130-150^{\circ}$, the second (low-energy) photon is
detected in the BGO ball or in the forward lead-scintillator wall
(Fig.~\ref{fig:kin}). This feature makes it possible to suppress
the $\pi^0$ photoproduction at backward angles $\theta_{cm} = 150
- 165^{\circ}$. At more forward angles one of the photons may
escape out from the detector through the backward hole.
Consequently, the background rejection
deteriorates dramatically.

\begin{figure}
\vspace*{0.6cm}
\epsfverbosetrue\epsfxsize=4.2cm\epsfysize=4.2cm\epsfbox{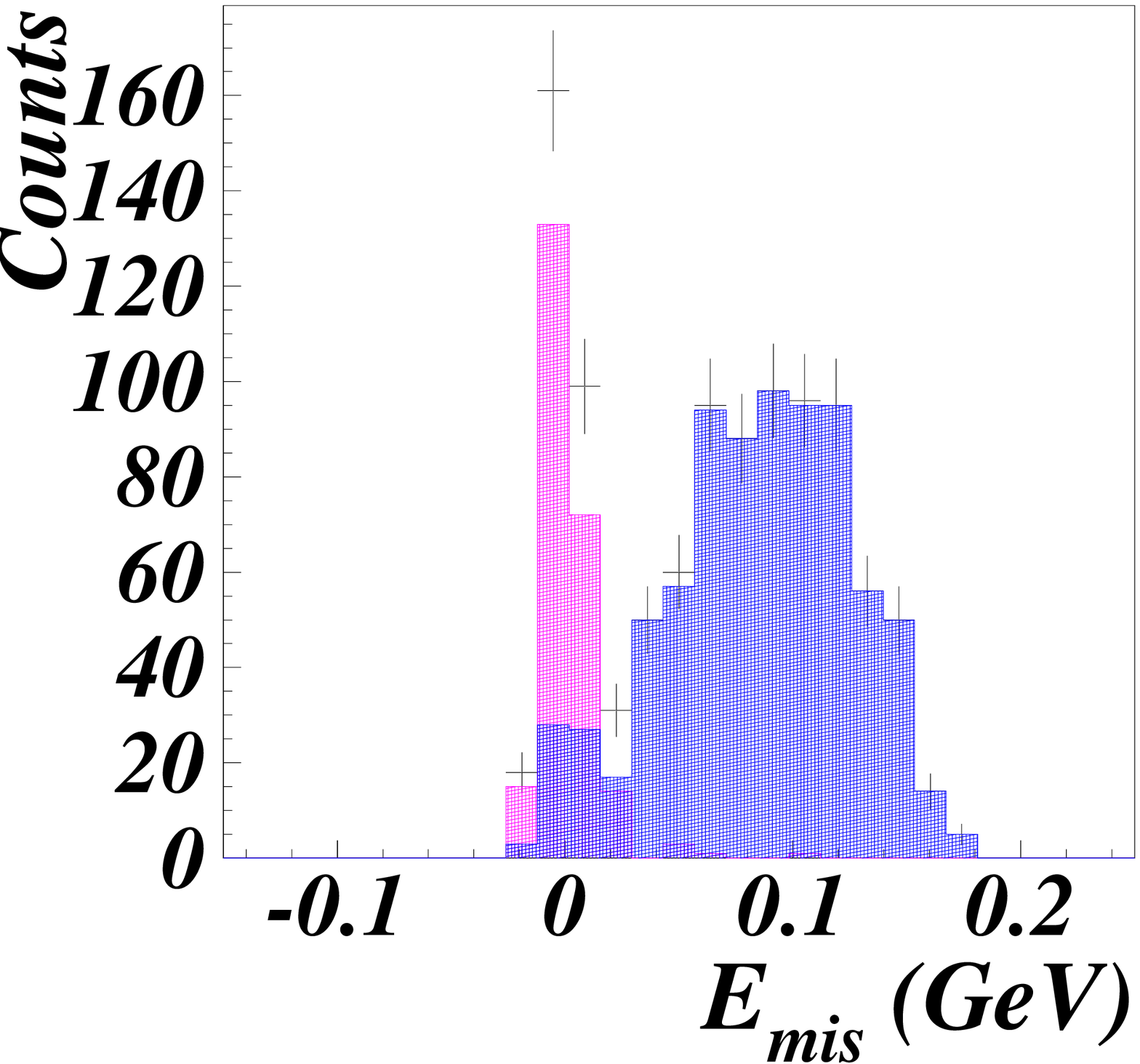}
\epsfverbosetrue\epsfxsize=4.2cm\epsfysize=4.2cm\epsfbox{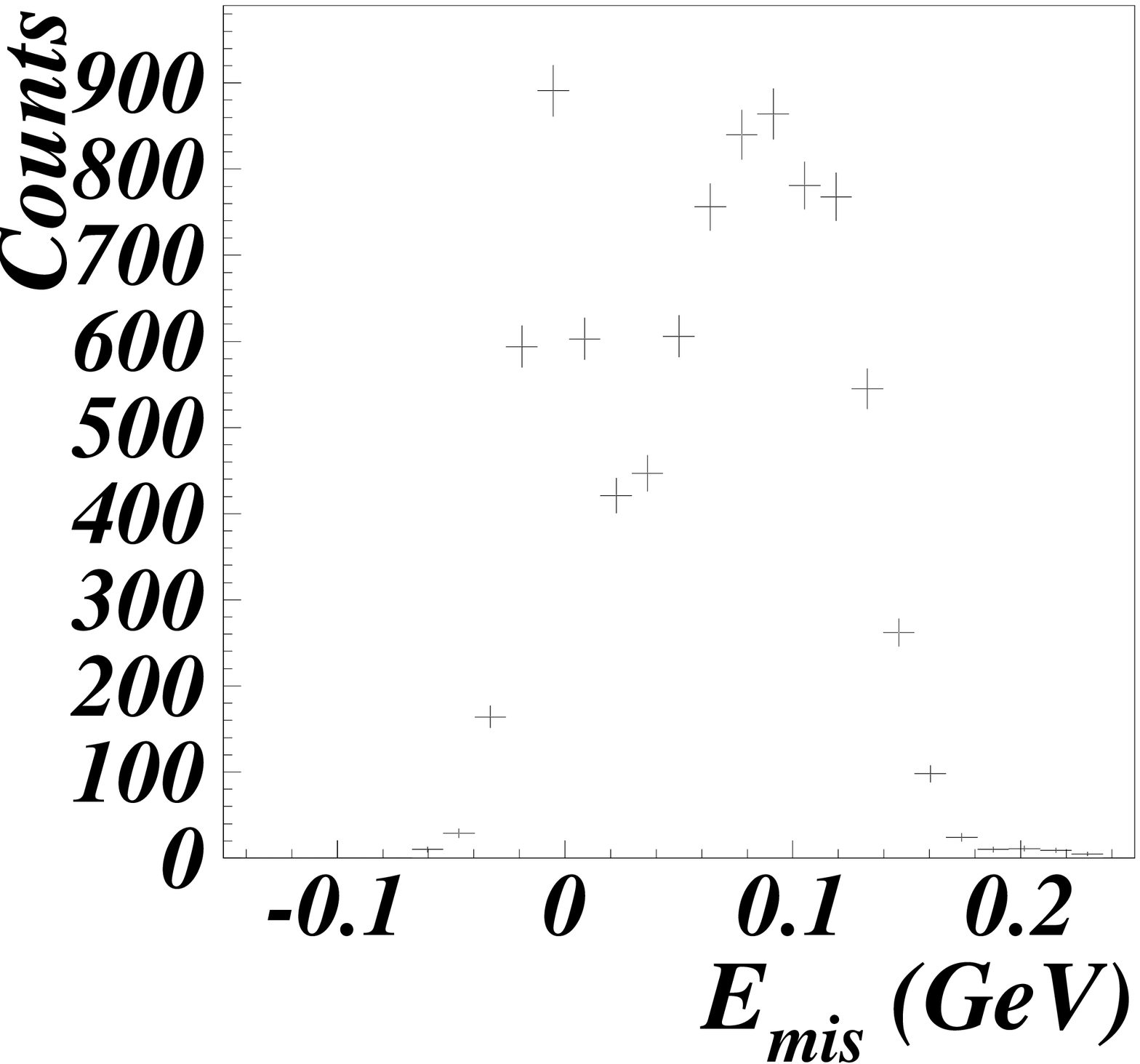}
\caption{On the left: Simulated spectrum of missing energy for a
free-proton target. The area colored in magenta shows Compton
events. The blue area corresponds to the photoproduction of
$\pi^0$s. On the right: Spectrum of missing energy measured in experiment with a free-proton target.}
\vspace*{-0.3cm} \label{fig:me} \vspace{-0.3cm}
\end{figure}

\begin{figure}
\vspace*{-1.9cm}
\epsfverbosetrue\epsfxsize=8.3cm\epsfysize=3.3cm\epsfbox{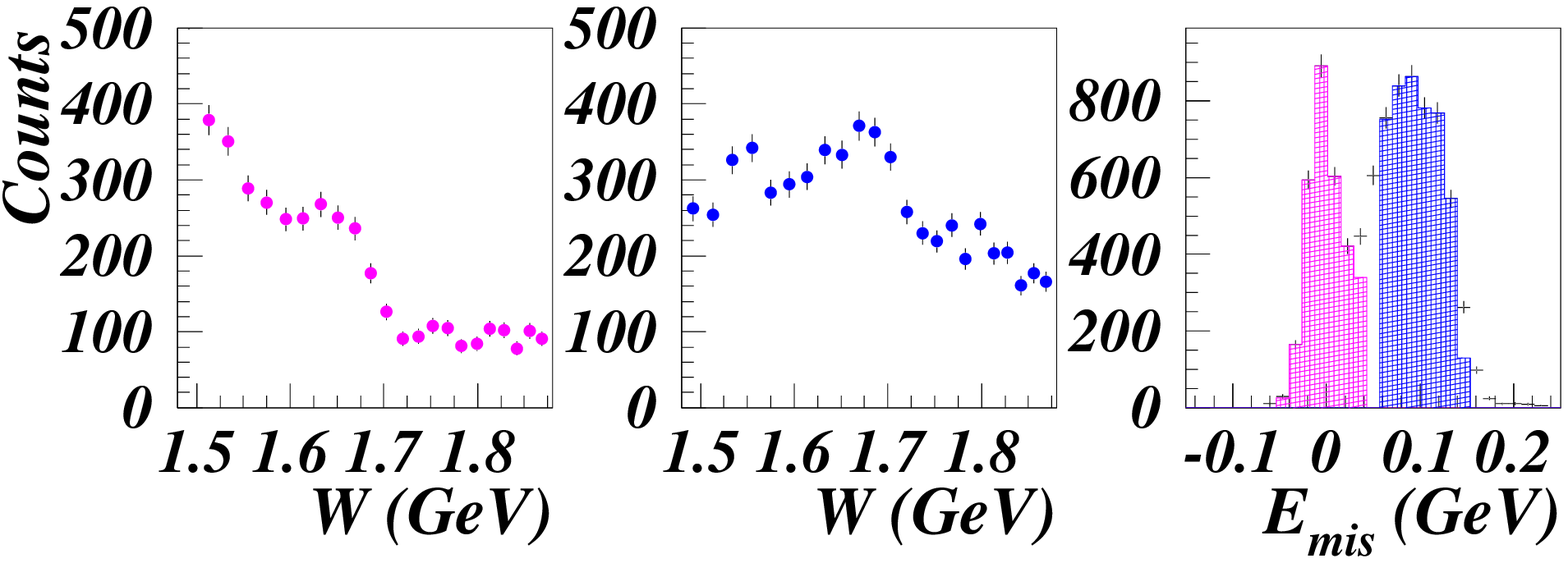}
\epsfverbosetrue\epsfxsize=8.3cm\epsfysize=3.3cm\epsfbox{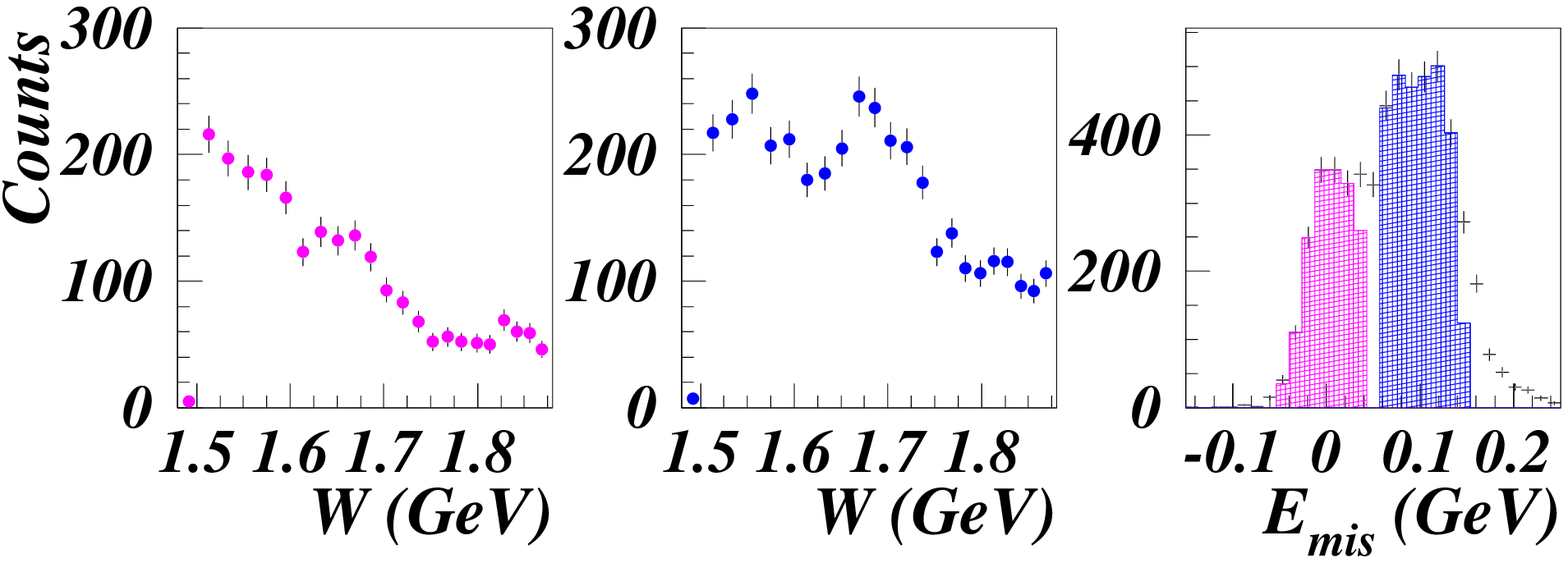}
\epsfverbosetrue\epsfxsize=8.3cm\epsfysize=3.3cm\epsfbox{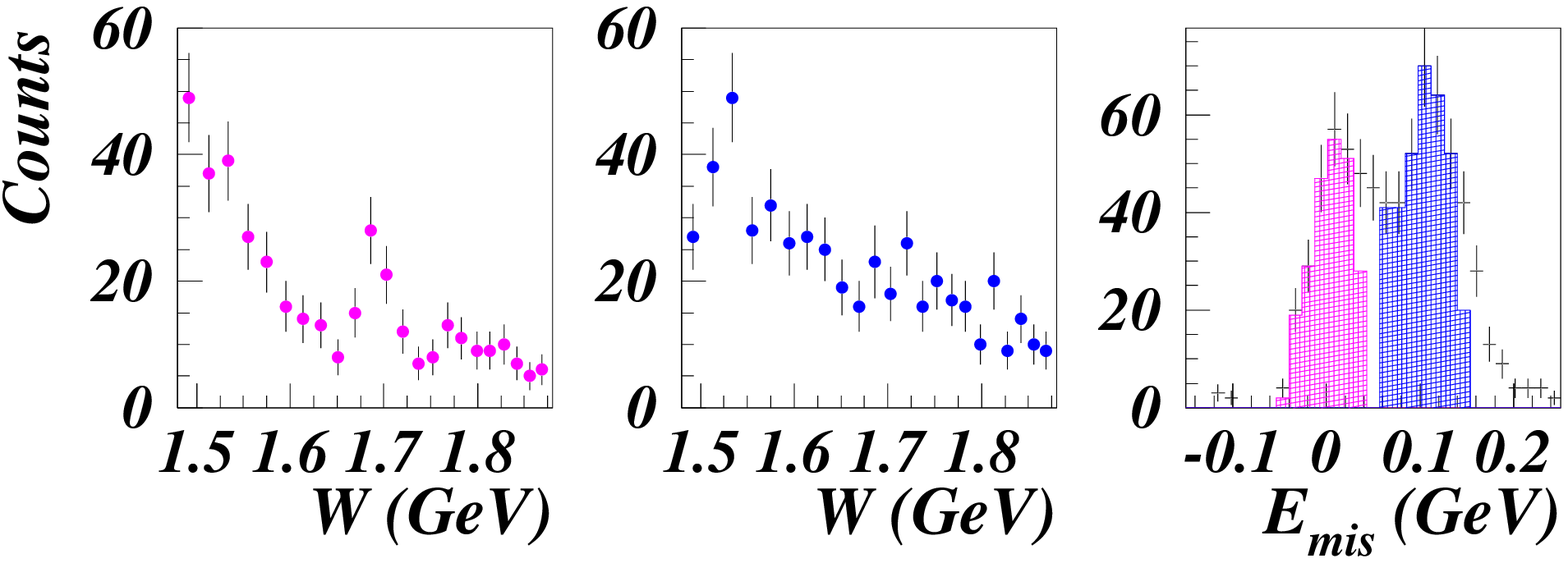}
\vspace*{2.3cm}
\caption{Experimental data obtained on the free proton (upper
row), quasi-free proton(middle row), and quasi-free neutron (lower
row). Right column: spectra of missing energy. Magenta and blue
areas indicate cuts used for the selection of Compton and $\pi^0$
events respectively. Middle column: W distributions of events
corresponding to blue areas in the missing-energy spectra ($\pi^0$
events). Left column: distribution of events corresponding to
magenta areas in the missing-energy spectra (dominance of Compton
events). } \vspace*{0.3cm} \label{fig:main} \vspace{-0.6cm}
\end{figure}

For the further selection of events the
missing energy $E_{mis}$ was employed
\begin{equation}
E_{mis}=E_{\gamma}-E_{{\gamma^\prime}}-T_{N}(\theta_{N}),
\end{equation}
\noindent where $E_{\gamma}$ denotes the energy of the incoming
photon, $E_{\gamma^\prime}$ is the energy of the scattered photon,
and $T_{N}(\theta_{N})$ is the kinetic energy of the recoil
neutron(proton).

The simulated spectrum of the missing energy for the free proton
is shown in the left panel of Fig.~\ref{fig:me}. $\pi^0$ events
form a wide distribution. Compton events generate a narrow peak
centered at $E_{mis}=0$. The  events in the region of this peak
mainly belong to Compton scattering. On the contrary, the cut $E_{mis}\geq
0.05$~GeV selects only $\pi^0$ events.
The right panel of Fig.~\ref{fig:me} shows the same spectrum measured with the free-proton target.
This spectrum is similar to the simulated one.

The right column of Fig.~\ref{fig:main} shows the missing energy
spectra corresponding to reactions on the free proton, (the first
row), the quasi-free proton (the second row), and the quasi-free
neutron (the third row). The data obtained on the quasi-free
nucleons are smeared by Fermi motion.

The left and central columns show the distributions of events
which correspond to the cuts $-0.05$~GeV
$\leq E_{mis} \leq 0.04$~GeV and $0.07$~GeV $\leq E_{mis} \leq
0.15$~GeV respectively. The first cut selects events around the
Compton peak. These events mostly correspond to Compton scattering
with some contamination of $\pi^0$ events. The second cut selects
mostly $\pi^0$ events.

The distributions of $\pi^0$ events obtained on the free and
quasi-free proton are similar and exhibit a wide bump near $W\sim
1.65$~GeV. This bump is well seen in the published data for this
reaction~\cite{pi0}. The Compton events on the proton indicate a
similar structure. This structure was also seen in  the previous
measurements~\cite{comp}. On the contrary, the distribution of
$\pi^0$ events on the neutron is flat. This observation is in
agreement with  the preliminary results from Crystal Ball/TAPS
\cite{kru1} and LNS Collaborations \cite{shim}.

The distribution of Compton events on the neutron (lower row, left
column of Fig.~\ref{fig:main}) reveals a narrow peak at $W\sim
1.685$~GeV. The peak is similar to that observed in the
$\eta$ photoproduction on the neutron.

\begin{figure}
\epsfverbosetrue\epsfxsize=2.8cm\epsfysize=3.4cm\epsfbox{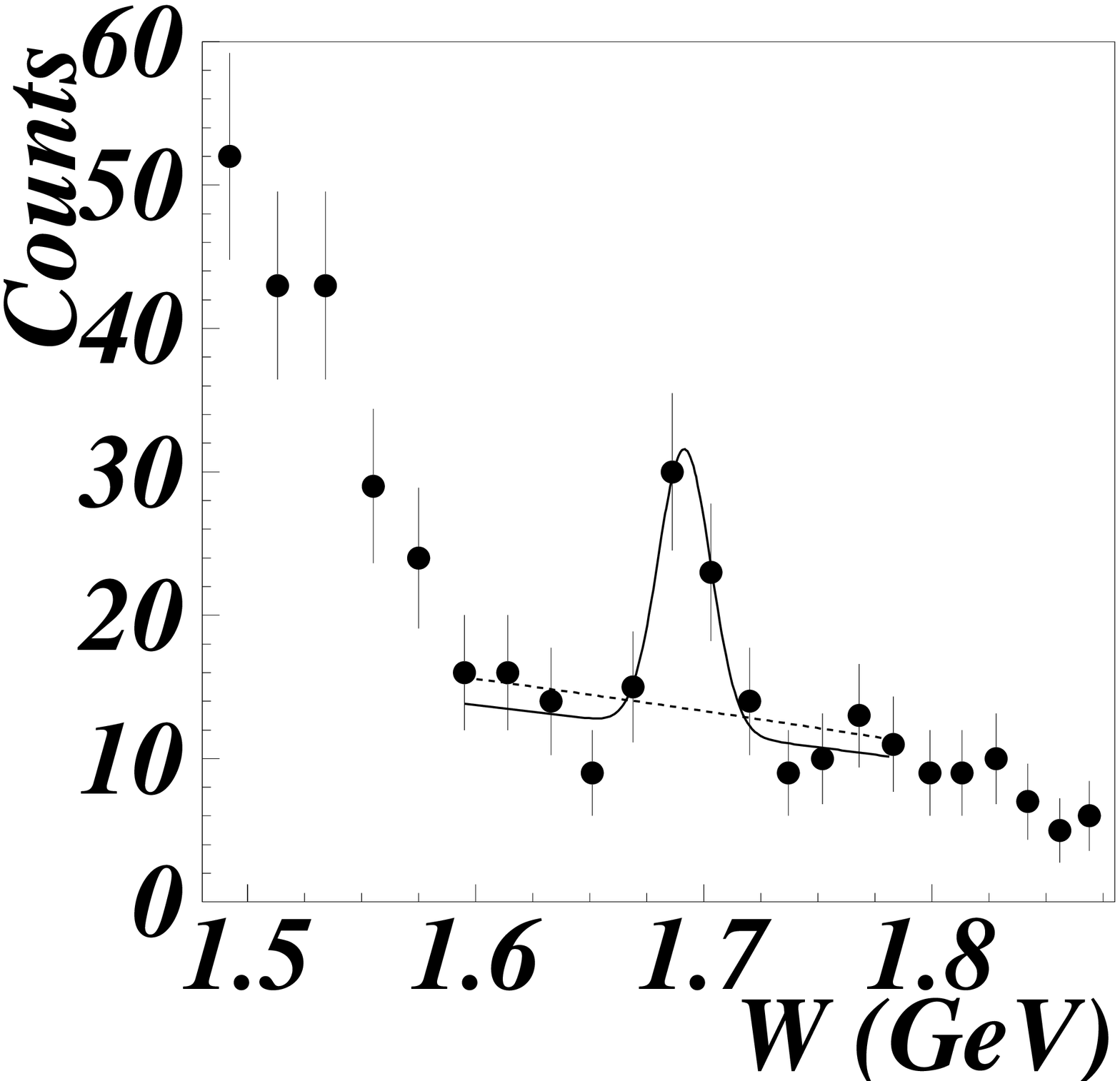}
\epsfverbosetrue\epsfxsize=2.8cm\epsfysize=3.4cm\epsfbox{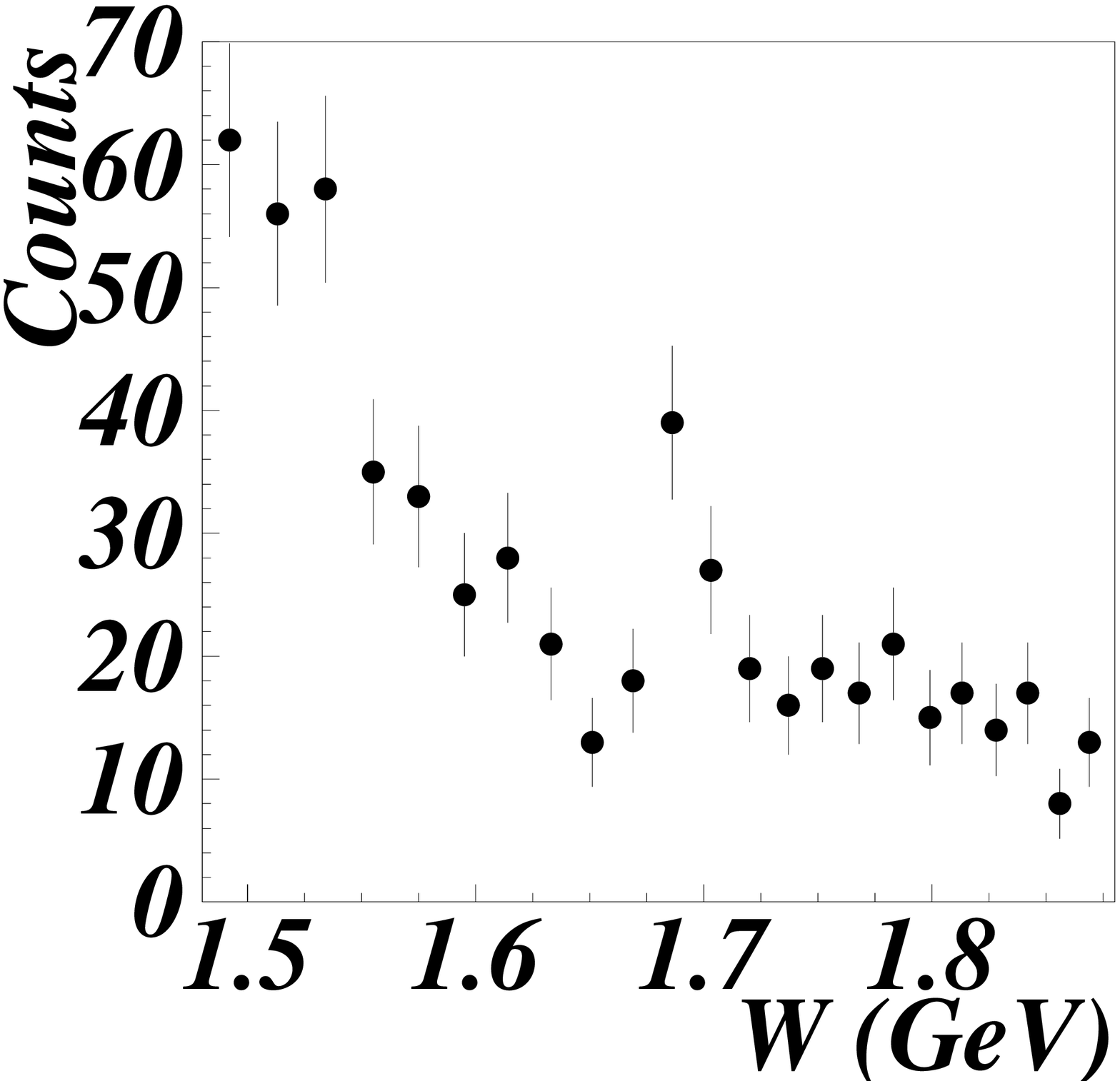}
\epsfverbosetrue\epsfxsize=2.8cm\epsfysize=3.4cm\epsfbox{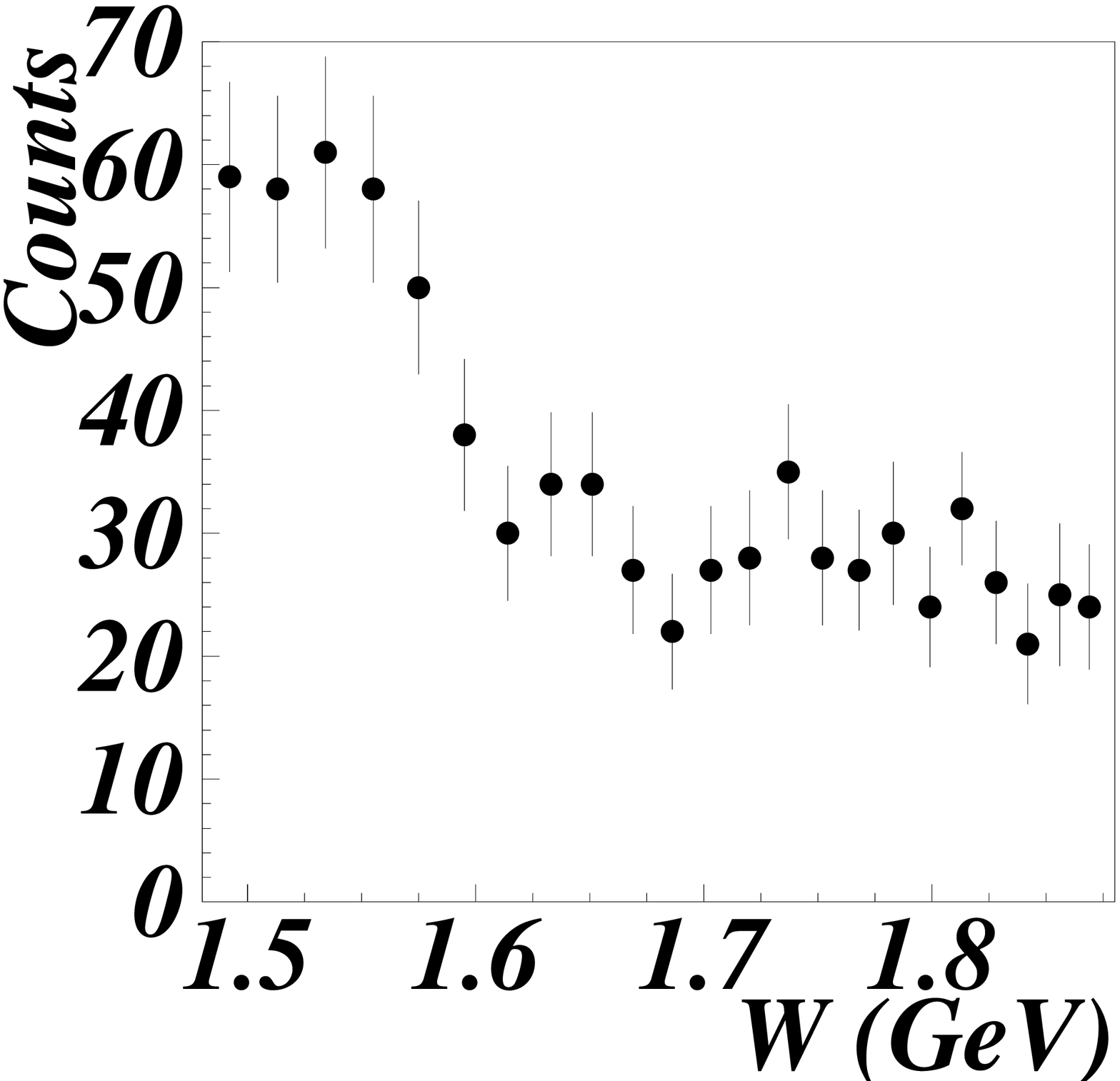}
\caption{Left panel: The W spectrum of events obtained with the cut on the missing energy
$-0.05\leq E_{mis} \leq 0.04 GeV$. Solid line indicates
the Gaussian-plus-2-order-polynomial fit. Dashed line corresponds to the 2-order-polynomial fit only. 
Middle panel: The W spectrum of events obtained with the cut $-0.1\leq E_{mis} \leq 0.075 GeV$.
Right panel: The simulated W spectrum obtained with the same cuts as in the left panel.}
\vspace*{-0.3cm}
\label{fig:fr} \vspace{0.3cm}
\end{figure}

In left panel of Fig.~\ref{fig:fr} the 2nd-order-polynomial (the background
hypothesis) fit for Compton events on the neutron in the interval
$W=1.585-1.888$~GeV is shown by the dashed line. The solid line in the same figure
shows the background-plus-Gaussian
fit. The $\chi^2$ of both fits are $3.7/6$ and $18.5/9$ respectively.
The log likelihood ratio of these two
hypotheses ($\sqrt{2\ln (L_{B+S}/L_B)}$) corresponds to the confidence level of $\sim 4.6\sigma$
The extracted peak position is $M=1686\pm 7_{stat}\pm 5_{syst}$~MeV.
and the \textit{r.m.s} is $\sigma\sim 12\pm 5 $~MeV ($\Gamma \approx 28 \pm 12 $ MeV).
The systematic uncertainty in the mass position is due
to the uncertainties in the calibration of the GRAAL tagger. 
 
The middle panel of  the Fig.~\ref{fig:fr} shows the similar distribution
obtained with the wider cut on the missing energy $-0.1\leq E_{mis} \leq 0.075 GeV$.
The contamination of the $\pi^0$ background is increased (espesially 
at the higher energies) while the peak at $W\sim 1.685$ GeV remains almost unaffected. 

The right panel of the Fig.~\ref{fig:fr} presents the simulated yield of events obtained 
with the same cuts as in the left panel of the same figure.
The event generator used in MC included a flat Compton cross section.
Neither of any peak appeared in the W spectrum of events. 

The observation of the narrow peak, its position and width, being considered 
togerther with the high-statistics results on the $\eta$ photoproduction
on the neutron~\cite{gra1,kru,kas,wert} and the beam asymmetry
data on the free proton~\cite{acta,jetp}, supports the existence of a narrow
nucleon $N^*(1685)$ resonance and challenges the
explanations~\cite{ani,skl} of the bump structure in the
quasi-free $\gamma n \to \eta n$ cross section in terms of the
interference of well-known resonances.

The assumption on the virtual sub-thershold $K\Lambda $ and $K\Sigma$ photopoduction~\cite{dor}
cannot be exluded. However, it requires the explanation why 
this effect occurs in the $\gamma n \to \eta n$ and $\gamma n \to \gamma n$ reactions 
and is not seen in $\gamma n \to \pi^0 n$.

The putative $N^*(1685)$ resonance is dominantly photoexcited on
the neutron whereas its photoexcitation on the proton is
suppressed. Such feature was suggested in Ref.~\cite{max} as the
benchmark signature of a resonance belonging to the
flavour $SU(3)$ antidecuplet of exotic baryons predicted by $\chi
QSM$~\cite{dia}. Interestingly, the mass, the narrow width, and
the isospin of $N^*(1685)$ are also in agreement with the
predictions for this member of the
antidecuplet~\cite{dia1,arndt,michal}.


The decisive identification of $N^*(1685)$, in particular its
definite association with the second member of the exotic antidecuplet,
requires further efforts and more experimental data. A critical
point is to determine the spin and the parity  of this state. It
is worthwhile to note that the fit of the beam asymmetry data for
the $\eta$ on the proton resulted in three possible quantum
numbers, namely $P_{11}$, or $P_{13}$, or
$D_{13}$~\cite{acta,jetp}.

In summary, we report the evidence for a narrow resonance
structure in the Compton scattering on the neutron. This structure
is quite similar to that observed in $\eta$ photoproduction on the
neutron. The combination of experimental observations suggests the
existence of a narrow nucleon resonance with unusual properties:
the mass $M\approx 1.685 $~GeV, the narrow width $\Gamma \leq 30$~MeV,
the much stronger photoexcitation on the neutron than on the
proton, and the suppressed branching ratio to $\pi N$ final
states.

It is our pleasure to thank the staff of the European Synchrotron
Radiation Facility (Grenoble, France) for the stable beam
operation during the  experimental runs. This work was
supported by Basic Science Research Program throgh the
National Research Foundation of Korea (NRF) funded by the
Ministry of Education, Science and Technology (grant 2010-0013430), 
and  by SFB/Transregio~16 (Germany). The work of M.V.P., I.A.P. and A.N.V.
is also supported by the grant 2010-1.5-508-005 of Russian
Ministry for Education and Research. The authors are grateful to
N. Sverdlova for her comments on the manuscript, and to Jiyoung Ha for 
the administrative support of this work.



\begin{thebibliography}{10}
\bibitem{an1} It is worth to noting that the authors of Ref.~\protect\cite{ann} arrived at a different conclusion. The detailed critique of the results from Ref.~\protect\cite{ann} is available in Ref.~\protect\cite{acta}. This critique remains unreplied.
\end{thebibliography}
\end{document}